\documentclass[aip,jcp,floatfix,preprint,citeautoscript]{revtex4-1}
\setcitestyle{super}

\usepackage{graphicx}
\usepackage{setspace}[1.5] 
\usepackage{amsmath, amsthm, amssymb}
\usepackage{xfrac}
\usepackage{color}

\newcommand{\etal}{\textit{et al.}~}
\newcommand{\s}{$\sim$}
\newcommand{\xc}{\textit{xc}~}
\newcommand{\kpoint}{\textbf{k}-point}

\begin{document}

\date{\today}

\title{How strongly do hydrogen and water molecules stick to carbon
  nanomaterials?}

\author{Yasmine S. Al-Hamdani}
\affiliation{Thomas Young Centre and
  London Centre for Nanotechnology, 17--19 Gordon Street, London, WC1H
  0AH, U.K.}
\affiliation{Department of Chemistry, University College
  London, 20 Gordon Street, London, WC1H 0AJ, U.K.}

\author{Dario Alf\`{e}}
\affiliation{Thomas Young Centre and London Centre for Nanotechnology,
  17--19 Gordon Street, London, WC1H 0AH, U.K.}
\affiliation{Department of Earth Sciences, University College London,
  Gower Street, London WC1E 6BT, U.K.}

\author{Angelos Michaelides} \email{angelos.michaelides@ucl.ac.uk}
\affiliation{Thomas Young Centre and London Centre for Nanotechnology,
  17--19 Gordon Street, London, WC1H 0AH, U.K.}
\affiliation{Department of Physics and Astronomy, University College
  London, 20 Gordon Street, London, WC1H 0AJ, U.K.}

\begin{abstract}
The interaction strength of molecular hydrogen and water to carbon
nanomaterials is relevant to, among many applications, hydrogen
storage, water treatment, and water flow. However, accurate
interaction energies for hydrogen and water with carbon nanotubes
(CNTs) remain scarce despite the importance of having reliable
benchmark data to inform experiments and to validate computational
models.  Here, benchmark fixed-node diffusion Monte Carlo (DMC)
interaction energies are provided for hydrogen and water monomers,
inside and outside a typical zigzag CNT. The DMC interaction energies
provide valuable insight into molecular interactions with CNTs in
general, and are also expected to be particularly relevant to gas
uptake studies on CNTs.  In addition, a selection of density
functional theory (DFT) exchange-correlation (\textit{xc}) functionals
and force field potentials that ought to be suitable for these
systems, is compared. An unexpected variation is found in the
performance of DFT van der Waals (vdW) models in particular. An
analysis of the peculiar discrepancy between different vdW models
indicates that medium-range correlation (at \textit{circa} 3 to 5 \AA)
plays a key role inside CNTs, and is poorly predicted by some vdW
models.  Using accurate reference information, this work reveals which
\textit{xc} functionals and force fields perform well for molecules
interacting with CNTs. The findings will be valuable to future work on
these and related systems, that involve molecules interacting with
low-dimensional systems.
\end{abstract}
\maketitle 

\section{Introduction}
Carbon nanotubes (CNTs) have been found to facilitate ultra-fast
diffusion\cite{holt2006fast,falk2010molecular,secchi2016massive,whitby2008enhanced,Majumder2005,ma2015water,su2015origin,michaelides2016nanoscience},
desalination\cite{tofighy2010salty,das2014carbon}, and water
treatment\cite{lee2015carbon}, and are also being developed into
biochemical
sensors\cite{volder2013,Kauffman2008,PengfeiQi2003,Dionisio2012,Liu2016sensor,Zanolli2011}.
The binding strength of molecules to carbon nanomaterials is
particularly important because it impacts upon macroscopic properties
such as the contact angle, slip length, and gas storage capacity of
nanomaterials, see \textit{e.g.}
Refs. \citenum{Lynn2013,das2014carbon,striolo2016carbon,muller2013purification,guo2015nanofluidic,bocquet2010nanofluidics}.
However, despite the thousands of papers on CNTs, the binding energy
of a single H$_2$ or water molecule on a CNT is still not well
established\cite{Ströbel2006781,liu2010hydrogen,Lei2016,Rubes2009cnt}.

Obtaining well defined experimental adsorption energies has always
been a challenge\cite{campbell_enthalpies_2013,wellendorff2015},
particularly so here, because of the difficulties in studying clean
surfaces of carbon nanomaterials and pure bundles of CNTs
experimentally. For instance, in the absence of benchmark adsorption
energies, H$_2$ was initially thought to adsorb on CNTs by as much as
\s200 meV and thus, CNTs were stipulated to be viable materials for
storing H$_2$
gas\cite{klechikov2015hydrogen,liu2010hydrogen}. However, more recent
estimates of the H$_2$ adsorption energy on CNTs suggest it is
considerably lower (\s50
meV)\cite{zuttel2002hydrogen,ye1999hydrogen}. Correspondingly, the
experimentally reported H$_2$ storage capacity of CNTs has decreased
in the past 20 years, falling from 14 wt.$\%$ to around 2
wt.$\%$\cite{liu2010hydrogen}.

Adsorption energies can be predicted from theory and this is
particularly useful given the scarcity of experimental reference
binding energies. However, it is essential to capture the weak
dispersion interactions that are prevalent in physisorption
systems. Modelling these interactions accurately is a formidable
challenge, especially in extended low dimensional
systems\cite{dobson2012calculation,ambrosetti2016wavelike,Ma11a,Voloshina_11,jenness2010benchmark,Jenness2009,Feller2000}
where the size of the system can pose an additional challenge. Since
various macroscopic properties hinge on the atomic-scale interactions,
even a small deviation in the adsorption energy can change the
predicted behavior of a system. For example, Hummer \etal have shown
that a range of adsorption energies and very small changes in the
water-carbon interaction can impact upon whether water enters a CNT or
not\cite{hummer2001water}. Therefore, it is important to have accurate
underlying models that provide reliable predictions.

The majority of computational studies focusing on either H$_2$ or
water on carbon nanomaterials use classical force fields with
Lennard-Jones (LJ) type potentials to model the intermolecular
interactions (see \textit{e.g.}
Refs.\citenum{falk2010molecular,fedorov2009thermoactivated,hummer2001water,striolo2016carbon}).
Density functional theory (DFT) is also seeing increasing application
for such systems (see \textit{e.g.} Refs.
\citenum{Lynn2013,zhao2002gas,agrawal2007ab,leenaerts2008adsorption,costanzo2012physisorption,silvestrelli2014including,fan2009prediction,han2004adsorption,horastani2013first,Peng2003}). However
predicted adsorption energies differ from one force field model to the
next, and the same is true for different DFT exchange-correlation
(\textit{xc}) functionals. It is not clear which of these methods give
more accurate results and only tentative comparisons can be made from
the literature, since different types of CNTs and adsorbate
configurations have been reported on. A number of force field studies
have relied on experimental adsorption energies of
H$_2$\cite{fedorov2009thermoactivated} and the contact angle of water
on graphite\cite{werder2003water} - a material which is physically
different to either graphene or CNTs. However, Leroy \etal have shown
that the ability to reproduce the correct adsorption energy between
water and the substrate in a force field leads to more accurate
results\cite{leroy2015parametrizing}, and therefore accurate reference
information is needed.

There are a number of ways to compute accurate adsorption energies and
here, benchmark interaction energies are provided from fixed-node
diffusion Monte Carlo (DMC) on an extended CNT for the first time.
DMC is explicitly correlated and accounts for exact exchange, thus it
is able to capture weak interactions that contain a significant
proportion of van der Waals (vdW) forces. Previously, Lei \etal
employed density fitted local coupled cluster with single, double, and
perturbative triple excitations (DF-LCCSD(T)) to compute interaction
energy curves for a water monomer with non-periodic, H-capped,
segments of CNTs of varying curvature\cite{Lei2016}.  This was an
incredibly impressive study, however, long-range charge density
fluctuations on the nanometre scale can impact upon the interactions
of low-dimensional systems like graphene and
CNTs\cite{dobson2012calculation,ambrosetti2016wavelike}. Accounting
for these long-range interactions requires one to go beyond localized
segments of such systems and instead, to use a periodic unit cell to
model an extended CNT. To this end, we have computed the physisorption
energy of both H$_2$ and water, inside and outside a CNT in a periodic
unit cell using DMC. The DMC reference interaction energies provide
insight into molecular adsorption on CNTs, and also indicate that the
uptake of H$_2$ and water is favored in the sub-nanometre CNT
considered here. We also compare the interaction energies with a
graphene substrate and draw similarities with adsorption on the
exterior of the CNT. In addition, direct comparison is made with some
new and some widely used \xc functionals and force fields.  We find
that a particular class of vdW \textit{xc} functionals overestimate
the interaction energy inside the CNT by up to twice as much. This
peculiar finding is considered more carefully, leading to some
important implications for molecular adsorption inside CNTs and vdW
methods.

\section{Methods}\label{meth}
The DFT calculations were performed with VASP
5.4.1\cite{vasp1,vasp2,vasp3,vasp4} with projector augmented wave
(PAW) potentials\cite{PAW_94,PAW_99}. There are countless \textit{xc}
functionals available in
DFT\cite{Becke2014,Yu2016pers,burke2012perspective} and it would not
be feasible to test all of them, hence only a few widely used or
relatively new \textit{xc} functionals have been chosen as part of
this study. The various \textit{xc} functionals tested include the
LDA\cite{LDA}, PBE\cite{PBE}, dispersion corrected PBE (D2\cite{D2},
D3\cite{D3a,D3b}, TSscs\cite{ts,scs,tsscsvasp}, and
MBD\cite{mbd,scs,buvcko2016many}) and vdW-inclusive functionals
(original vdW-DF\cite{vdwDF,revPBE}, optPBE-vdW\cite{vdw_opt10},
optB88-vdW\cite{vdw_opt10,B88}, optB86b-vdW\cite{vdw_opt11,B86},
vdW-DF2\cite{vdwDF2}, rev-vdW-DF2\cite{hamada2014van}). In the case of
the D3 correction, this is computed with the Becke and Johnson (BJ)
damping
function\cite{becke2005density,johnson2006post,becke2005exchange} and
with three-body Axilrod Teller contributions included. The revised
version of the Vydrov and Van Voorhis non-local correlation
functional, rVV10\cite{vv10,rvv10} is also included using Quantum
Espresso\cite{pwscf}. We have also tested the more recently developed
strongly constrained and appropriately normalized (SCAN) functional of
Sun \etal\cite{scan} This functional is expected to outperform PBE for
weakly interacting systems because it contains some non-local
correlation from constraints based on the non-bonded interaction of an
Ar dimer.

CNTs can vary in diameter, and can be either metallic or
semiconducting depending on their structure. The modelled CNT is
(10,0) in configuration, with a diameter of 7.8 \AA, and belongs to
the class of non-metallic zigzag CNTs. A CNT unit cell containing 80
carbon atoms was relaxed using a high 600 eV energy cut-off as
prescribed in VASP and PBE, PBE+TSscs, and vdW-DF2 functionals; the
resulting cell parameters differ by 0.7$\%$ at most. PBE+TSscs
predicted the nearest C-C bond length to the experimental C-C bond
length in graphite (1.421 \AA) at low
temperatures\cite{baskin1955lattice} and hence, the 8.58 \AA\ unit
cell length predicted by this functional along the CNT axis was chosen
for all further calculations. A unit cell length of $25$ \AA\ was used
along the other axes which allows for at least $\sim17$
\AA\ separation between periodic images of the CNT. The interaction
energy of water/CNT was tested against a larger CNT unit cell of 12.8
\AA\ length at the DFT level. The difference in interaction energies
was less than 3 meV indicating that the water is well separated from
its images. Water interaction energies were tested up to
$10\times1\times1$ \textbf{k}-points and convergence was reached
already with just the $\Gamma$-point (within 2 meV) and subsequently
used.

Graphene is a semi-metal for which a $(5\times5)$ unit cell was used
with a 15 \AA\ long vacuum between graphene sheets. Following a
convergence test on the number of \textbf{k}-points, a
$4\times4\times1$ \textbf{k}-point mesh was chosen. After careful
convergence tests for water/CNT and water/graphene interaction
energies, a plane-wave energy cut-off of 500 eV was applied for both
systems. Hard PAWs with 700 eV cut-off energy were also used to check
convergence and standard PAWs were converged to less than 0.2 meV for
the interaction energy of water on graphene.

The lowest energy geometries of water interacting with CNTs are not
entirely consistent in previous studies (varying by about 0.4 \AA)
which have mainly employed the LDA and
PBE\cite{zhao2002gas,sung2006ab,agrawal2007ab}. Here, vdW-DF2 and
PBE+TSscs were used to relax several starting configurations of water
and H$_2$, inside and outside the CNT, and on different sites above
graphene. The lowest energy orientations were found to be consistent
between PBE+TSscs and vdW-DF2 indicating that the choice of
\textit{xc} functional does not have a great impact on the adsorbate
geometry and vdW-DF2 relaxed structures were chosen for subsequent
DFT, force field and DMC calculations (see Fig.~\ref{figure1}). In
general, the potential energy surface is fairly smooth for graphene
and even more so for the CNT, and as such, we expect small variations
in the interaction energies for different configurations with other
methods.

Force field calculations were performed using
LAMMPS\cite{plimpton1995fast} with the TIP5P\cite{tip5p} and
SPC/E\cite{berendsen1987missing} models of water and
LCBOP\cite{los2003intrinsic} for the carbon substrates. The often used
Werder potential\cite{werder2003water} for carbon-water interaction
was tested along with recent carbon-water LJ type potentials that were
fit to coupled cluster data for water on
graphene\cite{perez2013anisotropy} and water on a H-capped segment of
CNT\cite{Lei2016}.

DMC calculations for CNT systems were performed using the CASINO
code\cite{casino} with the same cells and configurations as for the
DFT calculations. A similar approach to previous benchmark DMC
studies\cite{Ma11a,al-hamdani2,al-hamdani1} was used here.  A
plane-wave energy cut-off of 6800 eV was applied to the LDA
calculation of the trial wavefunctions in Quantum Espresso\cite{pwscf}
using the Trail and Needs pseudopotentials\cite{TN1,TN2} for all
atoms. The resulting wavefunctions were expanded in terms of B-splines
\cite{bsplines} using a grid spacing, $a={\pi}/{G_{max}}$, where
$G_{max}$ is the plane wave cutoff wavevector. A Jastrow factor with
up to three-body contributions was used to account for correlation and
optimised using variational Monte Carlo. 1-D periodicity was applied
along the CNT axis. For water/CNT systems, a time-step of 0.015
a.u. was used in DMC whilst employing the locality
approximation\cite{locapp} and 80,000 walkers for each
configuration. The DMC calculations were run until a stochastic error
of 10-15 meV was reached, producing a combined error of less than 20
meV in each interaction energy. A new implementation of CASINO
\cite{zen2016boosting} has been used for H$_2$ on CNT and graphene,
which allows a larger time-step (0.025 a.u.) to be used. A $3\times3$
unit cell of graphene was found to be large enough to avoid any
interaction between the periodic images of H$_2$ molecules at the DFT
level. In addition, \textbf{k}-point convergence was reached with 2
\textbf{k}-points at the DFT level. The total energy at each
\textbf{k}-point has equal weight in the total energy computed using 2
\textbf{k}-points. Trial wavefunctions were produced at each
\textbf{k}-point using Quantum Espresso, as prescribed for the CNT
systems. The resulting DMC energies at each \textbf{k}-point were
averaged to give a final interaction energy for H$_2$ on graphene.

The interaction energy of either water or H$_2$ on the carbon
substrates is defined as,
$E_{int} = E_{ads}^{tot} - E_{far}^{tot}$,
where $E_{ads}^{tot}$ is the total energy of the molecule/substrate
system in the interacting configurations shown in
Fig.~\ref{figure1}. $E_{far}^{tot}$ is the total energy of the
molecule/substrate system with the components separated by 12 \AA.  By
defining the interaction energy in this way, it has been shown that
size-consistency is maintained in the DMC calculation and the
time-step bias is also slightly reduced.\cite{zen2016boosting} This
definition of the interaction energy is used to report DMC, DFT and
force field results.

\section{Results}\label{res}
\subsection{Establishing accurate interaction energies using DMC}
The interaction energy of water and H$_2$ has been computed with DMC,
a selection of \textit{xc} functionals, and a few different classical
water-substrate force field models. Table \ref{table1} reports the
interaction energies for water at the CNT and in Table \ref{table2}
results for H$_2$ adsorption are reported. The reference DMC results
are discussed first, followed by the performance of the \textit{xc}
functionals, and finally some comments on the force field results are
presented.
\begin{figure}
\centering \includegraphics[width=0.85\textwidth]{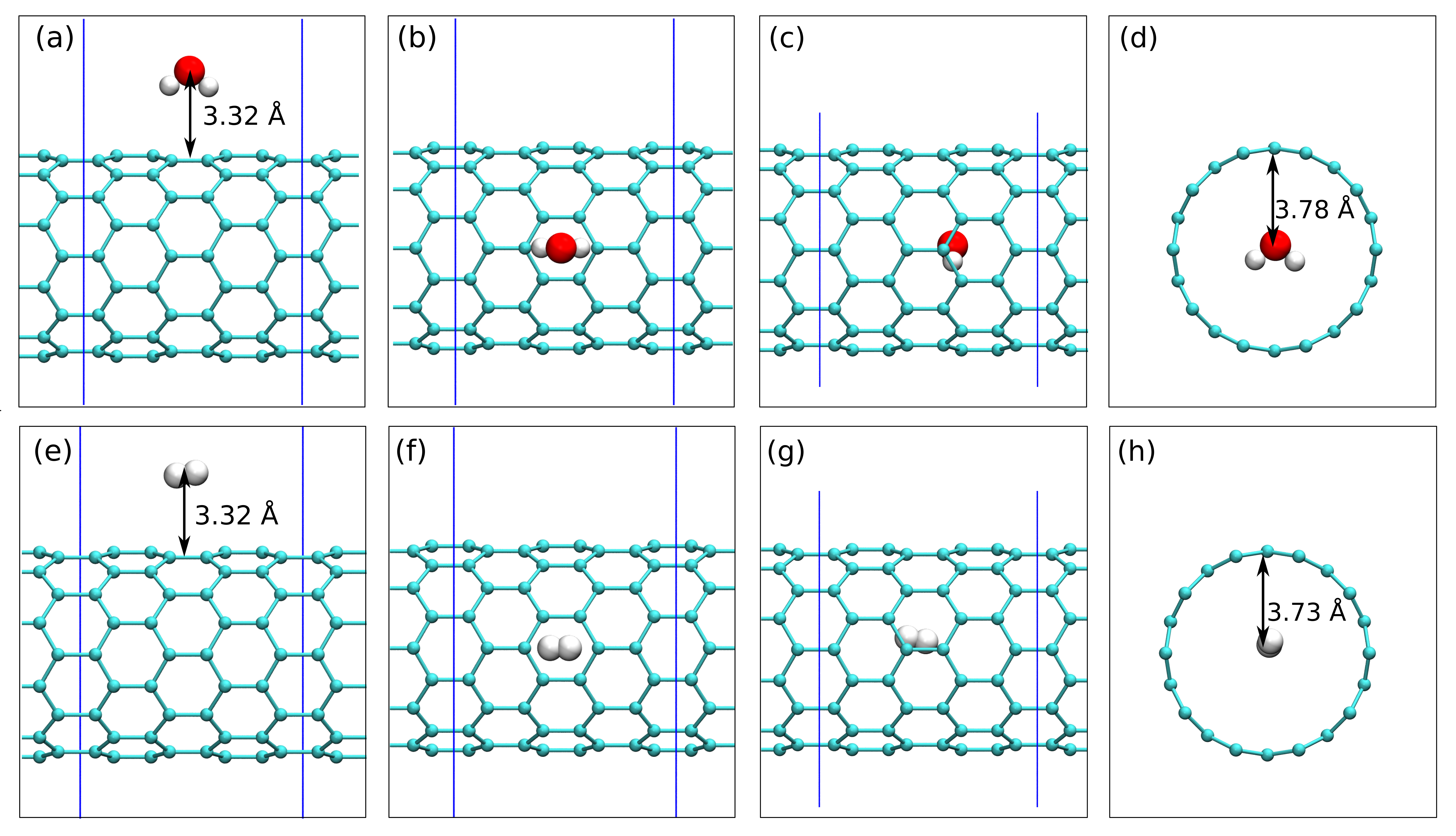}
\caption{Top panel: Unit cell of water outside (a and b) and inside (c
  and d) a CNT(10,0). Bottom panel: Unit cell of H$_2$ outside (e and
  f) and inside (g and h) a CNT(10,0). The unit cell is bounded by
  blue lines and contains 80 carbon atoms with a CNT diameter of 7.9
  \AA. Configurations have been obtained from DFT geometry relaxations
  with the vdW-DF2 functional.}
\label{figure1}
\end{figure}

Let us begin by discussing the DMC results for water and H$_2$. DMC
predicts that water physisorbs on the exterior of the CNT with an
interaction energy of $-80(\pm19)$ meV and on the interior with an
interaction energy of $-244(\pm17)$ meV. The DF-LCCSD(T) water
adsorption energies computed by Lei \etal for a H-capped CNT segment
with similar diameter\cite{Lei2016} are within \s20 meV of the DMC
reference energies reported here. Despite the many papers dedicated to
CNTs, experimental adsorption energies for water have not been
reported\cite{perez2013anisotropy,werder2003water} to the best of our
knowledge. As a result, water-carbon potentials for modelling CNTs
commonly rely on the water/graphite contact angle as a reference
instead\cite{werder2003water}.
Thus, theoretically computed adsorption energies of water/CNT serve as
references for experimental as well as computational studies focusing
on such systems.

In contrast to water, H$_2$ physisorbs more weakly: $-26(\pm10)$ meV
on the exterior of the CNT and $-115(\pm11)$ meV on the
interior. Similar adsorption energies were obtained for H$_2$ on a
metallic CNT in previous work by Rube{\v{s}} and Bludsk{\`y} using
coupled cluster corrected DFT\cite{Rubes2009cnt}. Our results show
that the H$_2$ interaction is \s50$\%$ weaker than water on each
substrate, likely because of the stronger electrostatic interaction
between the substrate and the permanent dipole of
water. Interestingly, estimated H$_2$ adsorption energies on CNTs from
temperature programmed desorption (TPD) experiments are reported
between 40 to 200
meV\cite{jones1997storage,ye1999hydrogen,shiraishi2003gas,darkrim2002review,hirscher2001hydrogen,hirscher2003carbon,panella2005hydrogen,zuttel2002hydrogen,cheng2001hydrogen}.
This large range has been attributed to different levels of purity of
CNT bundles used in experiments, and possible interference from metal
nanoparticles in the
samples\cite{hirscher2003carbon,hirscher2001hydrogen,darkrim2002review}. Our
DMC interaction energies suggest that H$_2$ adsorption energies on
pure CNTs should lie at the lower end of that range. The results also
imply that considerably higher measurements of H$_2$ adsorption
energies indicate the presence of impurities or defects in CNTs.
 
\begin{table}[ht]
\caption{{\label{table1}}Interaction energies (in meV/H$_2$O) of water
  outside the CNT, inside the CNT and on graphene. DMC energies are
  reported along with a selection of \textit{xc} functionals and force
  field models. Interaction energies that agree with DMC energies
  within the stochastic error are highlighted in bold.}
\begin{tabular}{lccc}
\hline
Method & external-CNT & internal-CNT & graphene                   \\ \hline
LDA\cite{LDA}                            &$-122$           &$\mathbf{-237}$ &$-124$         \\
PBE\cite{PBE}                            &$-26 $           &$-84 $          &$-21 $         \\
SCAN\cite{scan}                          &$\mathbf{-78}$   &$-203$          &$-84$          \\ \hline
\textit{Dispersion corrected xc functionals} & & &                \\
PBE+D2\cite{PBE,D2}                      &$-120$           &$-305$          &$-136$         \\
PBE+D3\cite{PBE,D3a,D3b}                 &$-113$           &$-293$          &$-126$         \\
PBE+TSscs\cite{PBE,ts,scs}               &$-137$           &$-327$          &$-158$         \\
PBE+MBD\cite{PBE,mbd,mbd14}              &$\mathbf{-99}$   &$-293$          &$-130$         \\
SCAN+D3\cite{D3a,D3b,scan,scand3}        &$-117$           &$-292$          &$-123$         \\ \hline
\textit{Dispersion inclusive xc functionals} & & &                \\
vdW-DF\cite{vdwDF,revPBE}                &$-109$           &$-458$          &$-130$         \\
optB88-vdW\cite{vdw_opt11,B86}           &$-123$           &$-457$          &$-152$         \\
optPBE-vdW\cite{vdw_opt10}               &$-137$           &$-506$          &$-169$         \\
optB86b-vdW\cite{vdw_opt11,B86}          &$-122$           &$-459$          &$-154$         \\
vdW-DF2\cite{vdwDF2}                     &$-108$           &$-397$          &$-129$         \\
rev-vdW-DF2\cite{hamada2014van}          &$\mathbf{-97}$   &$-365$          &$-119$         \\
rVV10 \cite{vv10,rvv10}                   &$-124$           &$-382$          &$-144$         \\ \hline
\textit{Force field methods}                 & & &                \\
Werder \etal\cite{werder2003water}       & $ -50$          & $-179$         &$ -63$         \\
Lei \etal\cite{Lei2016}                  & $ -123$         & $-360$         &$-156$         \\      
PHS\cite{perez2013anisotropy}            & $\mathbf{-99}$  & $-304$         &$-125$         \\ \hline           
\textbf{DMC}                             & $\mathbf{-80\pm19}$  & $\mathbf{-244\pm17}$ & $\mathbf{-70\pm10}^{Ma et al. 34}$\\ \hline
\end{tabular}
\end{table}
\begin{table}[ht]
\centering
\caption{Interaction energies of H$_2$ outside of the CNT, inside of
  the CNT, and on graphene in meV. A selection of \textit{xc}
  functionals and DMC energies are reported. Interaction energies that
  agree with DMC energies within the stochastic error are highlighted
  in bold.}
\label{table2}
\begin{tabular}{lccc}
\hline
Method & external-CNT & internal-CNT & graphene\\ \hline
LDA\cite{LDA}                            &$-60$          &${-96}$        &$-67$            \\
PBE\cite{PBE}                            &${-6}$         &$-22 $         &$-5$             \\
SCAN\cite{scan}                          &$\mathbf{-17}$ &$-50 $         &$\mathbf{-22}$   \\ \hline
\textit{Dispersion corrected xc functionals}  & & &               \\
PBE+D2\cite{PBE,D2}                      &$-48$          &$\mathbf{-117}$&$-59$            \\
PBE+D3\cite{PBE,D3a,D3b}                 &$-52$          &$-128$         &$-53$            \\
PBE+TSscs\cite{PBE,ts,scs}               &$-60$          &$-138$         &$-72$            \\
PBE+MBD\cite{PBE,mbd,mbd14}              &$-39$          &$\mathbf{-107}$&$-53$            \\
SCAN+D3\cite{D3a,D3b,scan,scand3}        &$-38$          &${-100}$       &$-43$            \\ \hline
\textit{Dispersion inclusive xc functionals}  & & &               \\
vdW-DF\cite{vdwDF,revPBE}                &$-59$          &$-230$         &$-77$            \\
optB88-vdW\cite{vdw_opt11,B86}           &$-59$          &$-216$         &$-75$            \\
optPBE-vdW\cite{vdw_opt10}               &$-74$          &$-253$         &$-94$            \\
optB86b-vdW\cite{vdw_opt11,B86}          &$-58$          &$-221$         &$-79$            \\
vdW-DF2\cite{vdwDF2}                     &$-55$          &$-181$         &$-69$            \\
rev-vdW-DF2\cite{hamada2014van}          &$-44$          &$-165$         &$-58$            \\
rVV10\cite{vv10,rvv10}                   &$-52$          &$-151$         &$-65$            \\ \hline
\textbf{DMC}   & $\mathbf{-26\pm10}$ & $\mathbf{-115\pm11}$ & $\mathbf{-24\pm11}$ \\ \hline
\end{tabular}
\end{table}

The DMC interaction energy of water with graphene has previously been
calculated to be $-70 (\pm10)$ meV\cite{Ma11a}. The interaction
energy of H$_2$ on graphene obtained here from DMC is $-24(\pm11)$
meV. The interaction energies of water on graphene and the exterior of
the CNT are very close in energy (within stochastic error). Likewise,
the DMC interaction energies for H$_2$ on the exterior of the CNT and
on graphene are within stochastic error.  The similar interaction
energies on graphene and outside the CNT suggest that the curvature of
this relatively small (10,0) nanotube has at most a modest impact on
the physisorption of small molecules on the exterior of the
CNT. Experimentally produced CNTs can have much larger diameters than
CNT(10,0)\cite{holt2006fast}, so it is likely that interaction
energies on those surfaces will be close to graphene.

Importantly, the DMC interaction energies inside the nanotube are
three times larger than those obtained outside the nanotube. This
relative difference between the interaction outside and inside of the
nanotube will have a large impact on molecules entering a
nanotube\cite{hummer2001water}. As such, it will be another important
aspect to consider when assessing the accuracy of various methods in
the following sections, starting with \textit{xc} functionals.

\subsection{Performance of \textit{xc} functionals: Challenge of internal interaction}\label{xcfuncs}
\begin{figure}[ht]
\centering \includegraphics[width=0.50\textwidth]{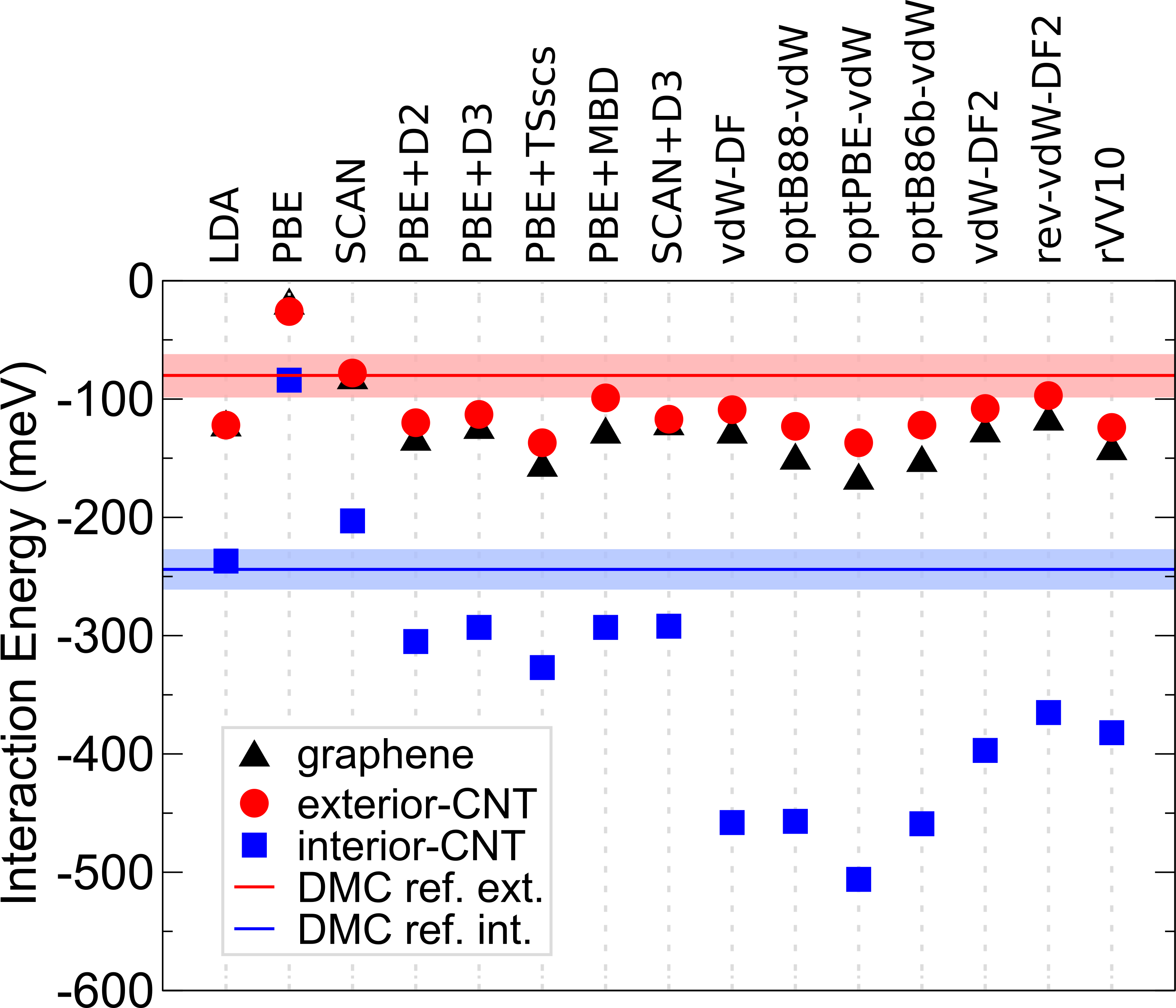}
\caption{Interaction energies of water inside (blue squares) and
  outside (red circles) the CNT with different \textit{xc} functionals
  and DMC. The DMC reference interaction energies are indicated by
  horizontal solid lines with the shaded area corresponding to the
  stochastic error.  The interaction energy of water on graphene with
  different \textit{xc} functionals is also shown using black
  triangles. All energies are in meV. }\label{figure2}
\end{figure}

\begin{figure}
\centering \includegraphics[width=0.50\textwidth]{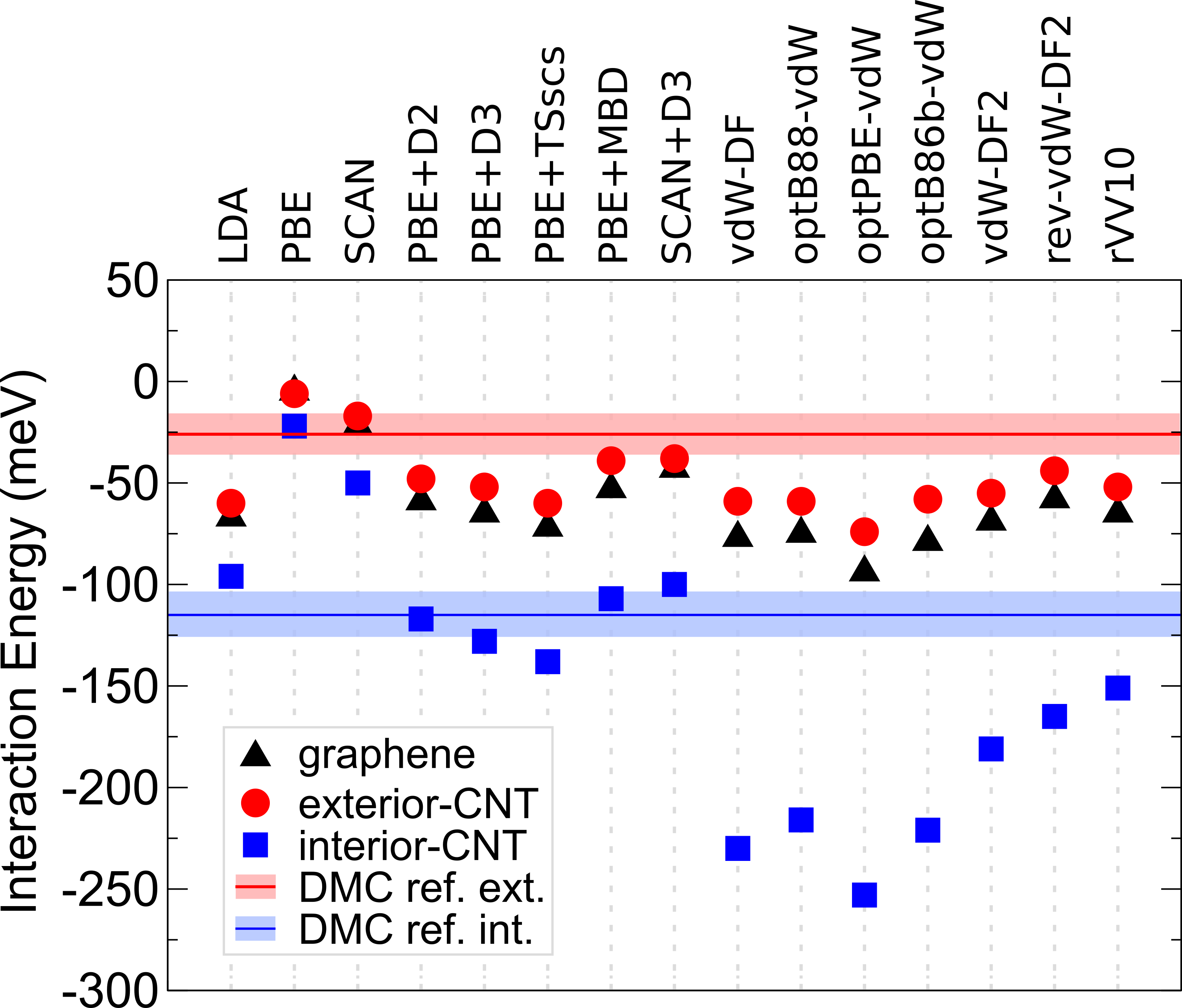}
\caption{Interaction energies of H$_2$ inside (blue squares) and
  outside (red circles) the CNT with different \textit{xc} functionals
  and DMC. The DMC reference interaction energies are indicated by
  horizontal solid lines with the shaded area corresponding to the
  stochastic error.  The interaction energy of H$_2$ on graphene with
  different \textit{xc} functionals is also shown using black triangles. All
  energies are in meV.}\label{figure3}
\end{figure}

With the reference DMC information we can assess the performance of a
selection of \textit{xc} functionals listed in Tables \ref{table1} and
\ref{table2}. We begin with the most commonly used functionals, the
LDA and PBE. The LDA only accounts for short-range correlation and yet
it overbinds both water and H$_2$ outside the CNT by up to 30 meV,
giving one of the worst performances for this configuration amongst
the \xc functionals considered. On the other hand, the LDA prediction
for water and H$_2$ adsorption inside the CNT, $-237$ and $-96$ meV,
respectively, is in close agreement with DMC. This fortuitous
performance of the LDA in physisorption systems is
well-known\cite{burke2012perspective,goodwater,al-hamdani2,williamson2002,galli2010}
and makes it difficult to draw physical insights from the LDA
predictions. We can see from Fig.~\ref{figure2} that PBE severely
underestimates the interaction energy of water on these low
dimensional carbon substrates wherein dispersion is a significant part
of the interaction. For H$_2$ adsorption PBE still underestimates the
interaction energy of the interior configuration but appears to
provide a closer estimate of the interaction energy for H$_2$ outside
of the CNT (see Fig.~\ref{figure3}).  The majority of previous DFT
studies on graphene and CNTs have used PBE and the LDA to study water
and
H$_2$\cite{zhao2002gas,agrawal2007ab,leenaerts2008adsorption,costanzo2012physisorption,silvestrelli2014including,han2004adsorption,horastani2013first}. The
reported water-substrate and H$_2$-substrate distances vary by up to
$\sim$1 \AA\ in the literature and involve CNTs with different
diameters and lengths. Even with these differences in mind, the
interaction energies in previous studies are within 30 meV of those
reported here for LDA and PBE.

More promising performance is seen for the recently developed SCAN
functional which predicts excellent physisorption energies for water
($-84$ meV) and H$_2$ ($-17$ meV) outside of the CNT. SCAN also
predicts a similar physisorption energy of water on graphene to DMC
(and RPA) from Ma \etal\cite{Ma11a} However SCAN slightly
underestimates the interaction energies by $\sim30$ meV for both
molecules inside the CNT. The underbinding results from the lack of
dispersion energy being taken into account.

There are two particularly common ways to account for dispersion
interactions in DFT \textit{xc} functionals. The first is by adding a
dispersion correction calculated from effective atomic dynamical
polarizabilities and includes the D2\cite{D2}, D3\cite{D3a,D3b},
TSscs\cite{ts,scs,tsscsvasp}, and MBD\cite{mbd,scs,buvcko2016many}
methods. Hence, this class of vdW methods is referred to as dispersion
corrected functionals. The second is based on the original vdW-DF from
Dion \etal\cite{vdwDF} in which two-body dispersion is calculated
based on charge densities and is seamlessly incorporated in an
\textit{xc} functional. This class of vdW methods is referred to as
vdW-DFs or dispersion inclusive functionals. Some vdW methods have
been shown to perform very well for weakly bound molecular systems (we
refer the reader to the reviews in Refs. \citenum{vdwpers,
  grimmerev,berland2015van,dobson2012calculation}); though not as well
for water adsorption on graphene\cite{Ma11a} and hexagonal boron
nitride\cite{al-hamdani2}.

For water/CNT and H$_2$/CNT, PBE+MBD and SCAN+D3 predict the best
interaction energies amongst the vdW methods tested here for both
exterior and interior adsorption. MBD takes into account beyond
two-body correlation interactions and is therefore able to capture
more effectively the dispersion that is present in the DMC reference
interaction energies. In the systems considered here, PBE+MBD predicts
the largest contribution from beyond two-body correlation interactions
for water inside the CNT, where it is $+26$ meV. Most of this
interaction energy arises from three-body interactions. Similarly, the
D3 correction includes up to three-body correlation interactions and
as we can see from Fig.~\ref{figure2} and Fig.~\ref{figure3} it also
performs well. The three-body correlation interaction predicted by the
D3 correction is $+34$ meV for water inside the CNT, in close
agreement with the MBD correction. The performance of these \xc
functionals is followed closely by the other dispersion corrected PBE
functionals.
  
For exterior adsorption of water and H$_2$ on the CNT, the vdW-DFs
perform similarly to the dispersion correction approaches:
over-binding by 20 to 40 meV compared to DMC. The exceptions are
vdW-DF2 and rev-vdW-DF2 which predict water interaction energies of
$-108$ and $-97$ meV, respectively. Rather strikingly, the vdW-DFs
predict significantly more pronounced interaction energies inside the
CNT, with up to a 250 meV overestimation by optPBE-vdW. That is twice
the DMC physisorption energy for water inside the CNT. In fact we see
two regimes emerge for vdW functionals based on internal interaction
energies from Fig.~\ref{figure2} and Fig.~\ref{figure3}. Such a stark
difference in the behavior of dispersion corrected DFT \textit{xc}
functionals and vdW-DFs is not often seen in other systems and raises
several questions which we address in the following section. However,
it is worth noting that all of the \textit{xc} functionals considered
here correctly predict that water adsorption is about twice as strong
as H$_2$ adsorption. Therefore, DFT \textit{xc} functionals are likely
to be useful methods for predicting the selectivity amongst molecules
for adsorption on CNTs. Moreover, all of the \textit{xc} functionals
with the exception of the LDA, correctly predict a \textit{circa}
threefold increase in the adsorption energy of molecules from outside
the CNT, to inside the CNT.

\subsection{Understanding the performance of DFT: The importance of  medium-range correlation}\label{analysis}
The DFT results in this study indicate that molecular adsorption on
CNTs is more accurately described by dispersion corrected \textit{xc}
functionals as opposed to including vdW interactions in a seamless,
though still approximate, manner. This is a somewhat unexpected
finding because such a clear-cut difference in interaction energies
between these two types of vdW functionals has not been observed
previously. The reader is referred to some notable reviews, for
example Refs. \citenum{vdwpers,grimmerev,berland2015van}, wherein
various vdW-DFs and dispersion corrected functionals have been
benchmarked on a number of weakly interacting systems, including the
S22 data set and H$_2$ adsorption on metal surfaces. In addition,
various assumptions made in developing these vdW functionals are
common to both types, and here we attempt to tease out the source of
the disagreement.

For vdW-inclusive functionals the charge density is immediately
brought into question since the dispersion contribution is calculated
using the densities. To address this possibility, the vdW-DF
interaction energy was calculated using the more localized
Hartree-Fock density of the water/CNT configurations. The reduction in
the interaction energy for the interior configuration of water is a
mere 11 meV, going from $-458$ meV to $-447$ meV. Hence, any
delocalization error that is present in the vdW-inclusive functionals
is not enough to explain the 100-200 meV overestimation seen here. For
completeness we also tested PBE0\cite{PBE0a,PBE0b} which is a hybrid
functional with a fraction of exact exchange. When combined with the
D3 dispersion correction, the resulting interaction energies of water
inside and outside the CNT are only 5 meV less than PBE+D3 energies.
Having established therefore, that exact exchange has very little
influence on the interaction of water with the CNT, we can proceed by
analysing the contribution from non-local correlation energy to the
interaction energies. Note that we use the term \textit{non-local
  correlation energy} interchangeably with \textit{dispersion energy},
to mean the long-range correlation interaction between electrons.
\begin{figure}[ht]
\centering \includegraphics[width=0.5\textwidth]{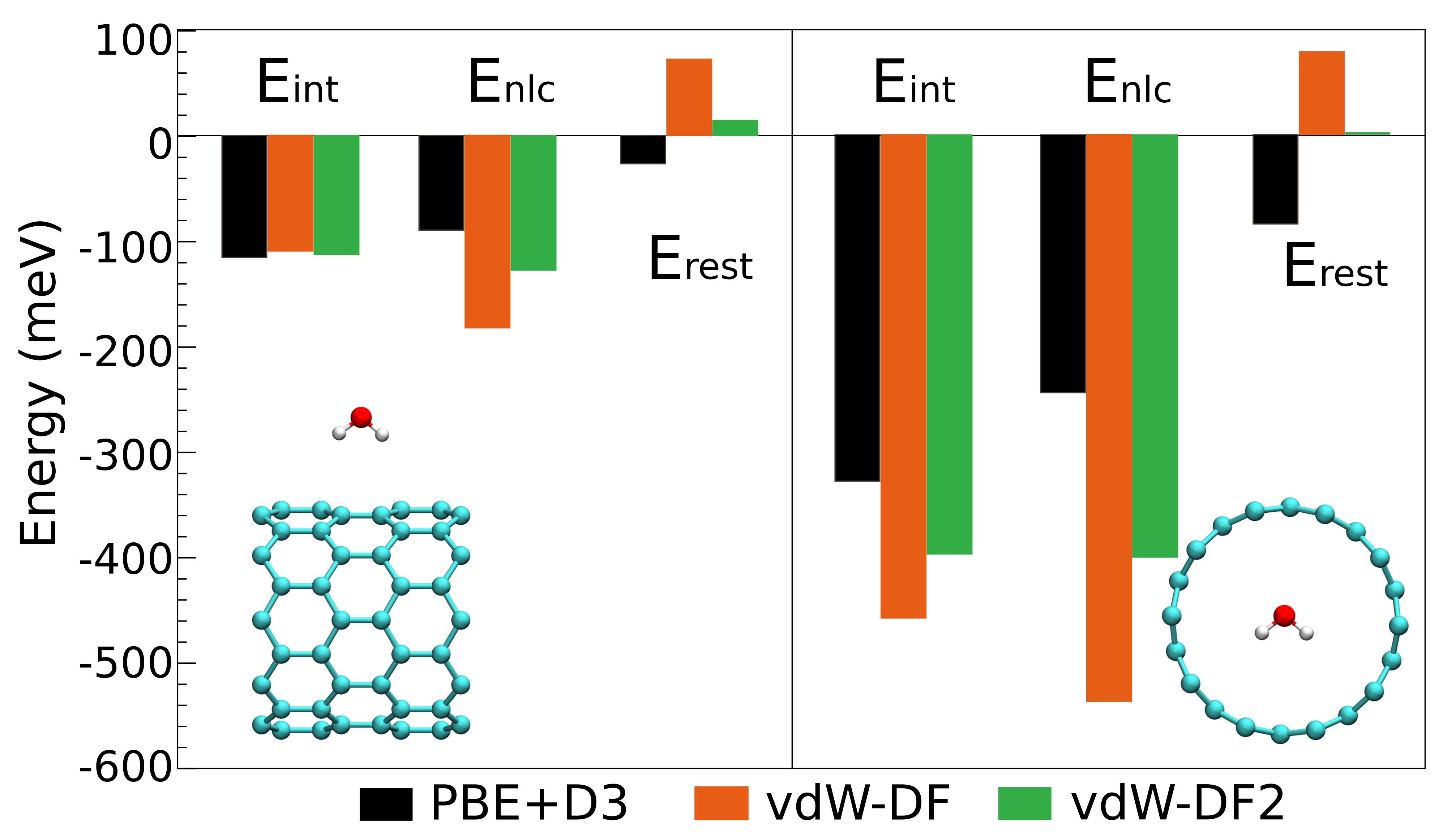}
\caption{Decomposition of the total interaction energy ($E_{int}$) for
  PBE+D3, vdW-DF and vdW-DF2. The contribution from non-local
  correlation energy ($E_{nlc}$) and all other interactions that are
  collectively referred to as $E_{rest}$, are
  shown. $E_{int}=E_{nlc}+E_{rest}$.}\label{figure4}
\end{figure}
\begin{figure}[ht]
\centering \includegraphics[width=0.5\textwidth]{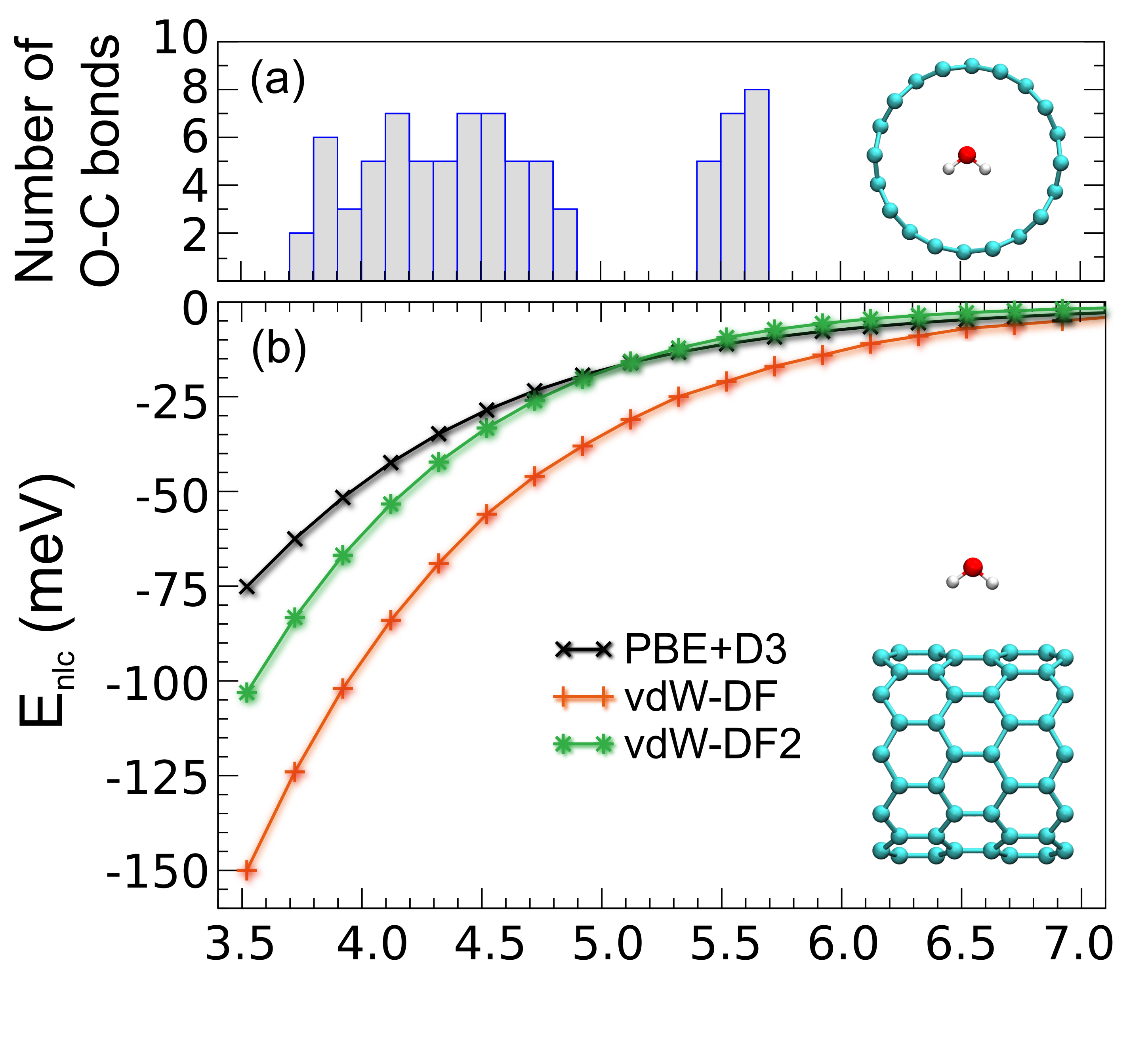}
\caption{(a) The barchart shows the number of oxgen-carbon bonds at
  different bond lengths for water inside of the CNT. (b) The
  non-local correlation energy curves are plotted from PBE+D3, vdW-DF
  and vdW-DF2, for water outside of the CNT: Single point calculations
  were used to compute the non-local correlation energy contribution
  ($E_{nlc}$) to the interaction of water outside the CNT at a series
  of oxygen-CNT distances. }\label{figure5}
\end{figure}

Fig.~\ref{figure4} shows an energy decomposition of the total
interaction energy of water inside and outside the CNT, for the
PBE+D3, vdW-DF, and vdW-DF2 functionals. The interaction energy is
decomposed into the contribution from non-local correlation energy
$E_{nlc}$, and all remaining components of the energy,
$E_{rest}$. Evidently from Fig.~\ref{figure4}, the contribution from
$E_{nlc}$ in the vdW-DFs is much larger than with the D3
correction. However, when water is outside the CNT, the larger
$E_{nlc}$ in the vdW-DFs is compensated by a repulsive interaction
from $E_{rest}$. As a result, the three functionals predict almost the
same interaction energy for water outside the CNT. On the other hand,
the contribution from $E_{rest}$ in the vdW-DFs is much the same
inside the CNT as it is outside (compare the left and right panel of
Fig.~\ref{figure4}). Whereas, there is a threefold increase in
$E_{nlc}$ from water outside the CNT to inside, and this increased
attraction inside the CNT is clearly not compensated by $E_{rest}$ in
the vdW-DFs. In other words, for molecules outside of the CNT, the
overestimation of non-local correlation interaction by vdW-DFs is
cancelled out by more repulsion in $E_{rest}$. This compensating
effect is not present for molecules inside the CNT.  The compensating
effect in the dispersion inclusive functionals is present by
design\cite{vdwDF, vdwDF2, vdw_opt10, vdw_opt11} to help their
accuracy on relatively small molecular dimers.

So why is water inside the CNT a particular problem for the dispersion
inclusive methods? To answer this, we look more closely at the
dispersion energy as a function of water-CNT distance for water
outside the CNT, and compare this with the oxygen-carbon distances for
water inside the CNT. This has been done by computing the interaction
energy curve of water outside the CNT with PBE+D3, vdW-DF, and
vdW-DF2, and extracting the contribution from dispersion energy
(\textit{i.e.}, $E_{nlc}$) at each point along the curves. The total
interaction energy curves can be found in the Supporting Information
(SI), but here we simply comment that the interaction energy curves
for water outside the CNT are very similar to that of water on
graphene.

In Fig.~\ref{figure5} the dispersion energy curves for water outside
the CNT can be seen to vary significantly between the three
functionals. As mentioned already, there is a pronounced repulsive
interaction in the vdW-DFs that alleviates the large non-local
correlation energy for water outside the CNT, but crucially, not for
water inside the CNT. Comparing these dispersion energy curves with
the frequency of oxygen-carbon bonds at a given distance for water
inside the CNT in Fig.~\ref{figure5}(a), we see that the majority of
oxygen-carbon bonds inside the CNT lie within 3.5 to 5.0 \AA. At these
distances the dispersion energy is particularly large in the vdW-DFs
compared to PBE+D3. In the absence of an adequate repulsive
interaction (as illustrated in Fig.~\ref{figure4}), the total
interactions at these medium-range distances are poorly described by
the vdW inclusive functionals. This could be interpreted as too much
correlation energy at medium-range distances or equally as not enough
repulsive interaction to compensate for it. Note that this
medium-range correlation regime refers to atomic separations larger
than bonding distances (a few {\AA}ngstroms) and closer than the
long-range limit where the interaction reaches the $1/r^6$ limit (\s10
\AA).

The reasonably good performance of dispersion corrected SCAN and PBE
suggests these describe the medium-range interactions better. This is
possibly due to the use of damping functions\cite{vdwreview,D3b}, that
are used to adjust the short-range behavior of the dispersion
correction with respect to the underlying \textit{xc} functional
empirically. In this way, damping functions directly affect the
medium-range interactions in the dispersion corrected functionals that
we have tested.

Although we have not come across any studies showing or discussing two
distinct regimes for the performance of vdW-DFs and dispersion
corrected functionals, there are indications of this finding in
previously computed interaction energy curves. In particular, the
ordering of some \textit{xc} functionals at medium-range distances in
the interaction energy curves of weakly interacting complexes in
Refs. \citenum{ganesh_binding_2014,klimes2011towards,Hamada2012,ruiz2012density,Graziano2012,
  al-hamdani2}, closely match the order we see in Figs.~\ref{figure2}
and~\ref{figure3}.

\begin{figure}[ht]
\centering \includegraphics[width=0.45\textwidth]{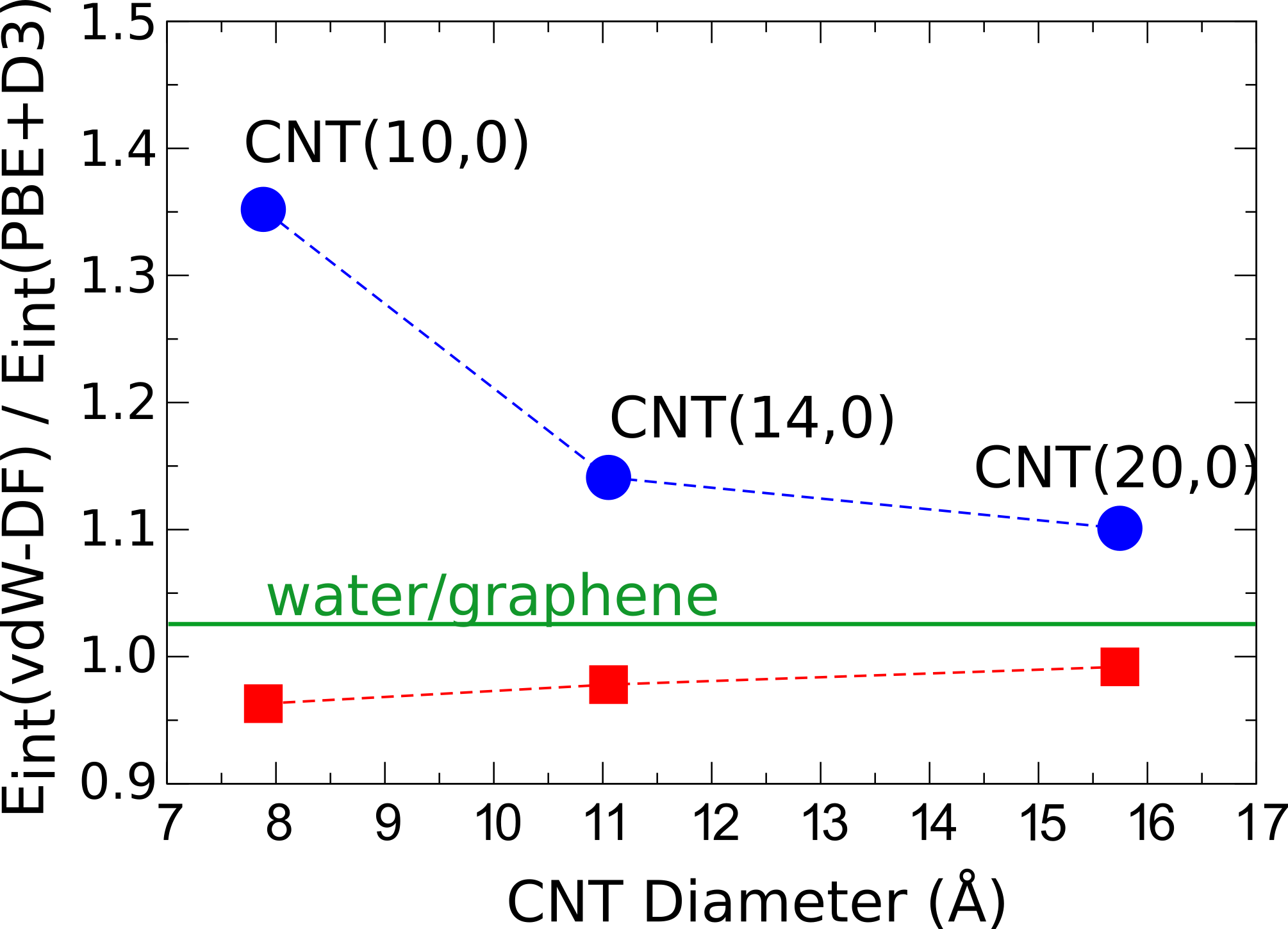}
\caption{Ratio of vdW-DF to PBE+D3 interaction energies for water
  inside (blue squares) and outside (red circles) of CNTs with
  increasing diameters: CNT(10,0), CNT(14,0) and CNT(20,0). The
  water/graphene ratio is indicated by the solid green
  line.}\label{figure6}
\end{figure}

The importance of medium-range correlation can also be seen by
comparing the geometry optimized interaction energies for water inside
and outside of CNTs with different diameters in
Fig.~\ref{figure6}. Water interaction energies outside the CNT show
less than $7\%$ deviation between PBE+D3 and vdW-DF across all three
CNT diameters, shown in Fig.~\ref{figure6}. Whereas for water inside
the CNT, the interaction energy difference PBE+D3 and vdW-DF increases
rapidly from $9\%$ to $30\%$ as the CNT diameter decreases. For larger
CNT diameters, there are fewer oxygen-carbon bonds at medium-range
distances for water inside the CNTs. The corresponding radial
distribution functions between oxygen and carbon, $g_{OC}$, can be
found in Fig. 2 of the SI. This suggests that vdW-DF begins to
overestimate the dispersion interactions more than PBE+D3 for CNTs
with diameters less than $\sim10$ \AA. We expect this to be the case
for all the other vdW-DFs tested in this paper, as well as the rVV10
functional.

\subsection{Reliable water-carbon potentials for water/CNT?}\label{ffs}
The DMC simulations are also useful in helping to evaluate how
standard force field models for the water-carbon interaction
perform. Three LJ type force fields for the water-carbon substrate
interaction have been tested here, referred to as: Werder
\etal\cite{werder2003water}, Lei \etal\cite{Lei2016}, and PHS
(P{\'e}rez-Hern{\'a}ndez and Schmidt)\cite{perez2013anisotropy}. The
potential by Werder \etal is one of the most commonly used for
water/carbon systems and was designed to reproduce experimental water
contact angles on graphite\cite{werder2003water}. In this potential
only the C-O interaction is defined ($\epsilon_\text{CO}=4.549$ meV
and $\sigma_\text{CO}=3.19$ \AA) and it was obtained by tuning
$\epsilon_\text{CO}$ until an experimental water/graphite contact
angle was reproduced with the SPC/E model of water. It can be seen
from Table \ref{table1}, that this interaction potential leads to an
underestimation in the interaction energy of water especially inside
the CNT, where it is at least 40 meV (20$\%$).

Lei \etal have suggested a few different water-carbon potentials by
manually fitting interaction parameters to DF-LCCSD(T) interaction
energy curves for water with H-capped segments of CNTs. It is
recognised therein that water adsorption inside and outside the
nanotube is not accurately predicted by any single set of
parameters. We have chosen one that includes C-H interaction
parameters as well ($\epsilon_\text{CH}=4.457$ meV and
$\sigma_\text{CH}=2.80$ \AA)\cite{Lei2016}. Using TIP5P for the water
force field as prescribed, there is a threefold increase of the water
interaction energy from exterior ($-123$ meV) to the interior ($-360$
meV) of the CNT, in agreement with the ratio from DMC. However, the
interaction energies are overestimated outside (by \s20$\%$) and
inside the CNT (by \s40$\%$). It is worth noting however that the
orientation of water in the DF-LCCSD(T) calculations is different to
the configuration studied here; with the H atoms of water parallel to
the length of the CNT instead of perpendicular as shown in
Fig.~\ref{figure1}.

Another LJ type water-carbon potential based on the CCSD(T)
water-graphene adsorption energy and the TIP5P model of water has been
proposed by P{\'e}rez-Hern{\'a}ndez and
Schmidt\cite{perez2013anisotropy}. This PHS model was obtained by
reproducing the CCSD(T) interaction energy of water in the up and down
configurations on a 58 carbon atom segment of
graphene\cite{Voloshina_11}. Orientation dependence is therefore built
in by defining C-H as well as C-O interactions for water.  From Table
\ref{table1} it can be seen that this potential performs very well,
predicting $-99$ meV for water outside of the CNT, which is within the
stochastic error of the DMC reference. In addition, for water inside
the CNT the PHS force field performs as well as the dispersion
corrected functionals (see Table \ref{table1}).

The sensitivity of the force fields to the form of parametrization is
clear from the varying performance of the three force field models
considered here. With the DMC reference interaction energies of water
on the CNT, we can see that the PHS force field is particularly good
for these systems -- performing on par with dispersion corrected
\textit{xc} functionals. As demonstrated, the DMC reference
interaction energies in this study could be used to determine the
accuracy of future force field adsorption studies on these systems.

\section{Discussion}\label{disc}
The benchmark DMC energies reported in this paper are the first,
explicitly correlated and exact exchange, interaction energies for
water and H$_2$ on an extended CNT and are also intended to serve as
references for other methods. Additional insight is given on the DMC
results in this section and the significant overestimation by vdW-DFs
for adsorption inside the CNT is addressed.  We first comment on the
appropriateness of the DMC method for these systems in the context of
other benchmark methods, and we make an estimate of finite size
effects in the DMC energies.  Later, we comment on the findings in the
context of other types of nanotubes namely, metallic CNTs and
insulating boron nitride nanotubes (BNNTs).

Let us first consider the suitability of DMC for interaction of
molecules with the CNT(10,0) that is considered here.  The DMC
calculations in this study used a single-determinant approach. This is
expected to be sufficient since multi-reference character is unlikely
given that the band gap of CNT(10,0) is \s1 eV even at the GGA
level\cite{Deible2015,Cohen2008,Yu2016}. Furthermore, an important and
challenging feature of CNTs that needs to be accounted for is their
extended and delocalized nature. To this end, DMC can be efficiently
used with periodic boundary conditions and as a result, it is free of
localization approximations in the charge density and
polarizability. On the contrary, such approximations are inherent in
non-periodic calculations using CCSD(T). Using unit cells with
periodic boundary conditions however, leads to finite size effects in
DMC that merit further comment.

The main source of finite size effects relevant to the CNT studied
here with DMC is the long-range nature of Coulomb interactions. Such
long-range Coulomb interactions can extend to the nanometre
scale\cite{ambrosetti2016wavelike} and are prevalent in
low-dimensional extended materials with small band
gaps\cite{white2008enhanced,ambrosetti2016wavelike}. Capturing
long-range interactions at the nanometre scale requires unit cells
that extend to the same lengths as the interactions, \textit{i.e.} a
few nanometres. The unit cell used in this work contains 80 carbon
atoms and is 8.58 \AA\ along the CNT axis. Although DMC provides a
many-body solution for this relatively large system, larger unit cells
become increasingly prohibitive. Instead, finite size effects can be
estimated in the unit cell used here by invoking the MBD correction at
the DFT level.  Unlike DMC, the MBD correction is computationally
inexpensive and can therefore be used in large unit cells to capture
the contribution from long-range Coulomb interactions.

We find that the MBD correction to PBE is converged with a
\kpoint\ mesh of $2\times1\times1$ which is equivalent to doubling the
unit cell along the length of the CNT. The MBD correction increases
the water interaction energy by 12 meV outside the CNT and by 16 meV
inside the CNT. For the H$_2$ interaction the MBD correction is 5-8
meV only. These corrections are applicable to all calculations with a
periodic unit cell, including DMC, and should be taken as the finite
size error corrections. Shifting the DMC reference energies in this
manner increases the interactions, but since the corrections are
relatively small they remain within the stochastic error bars of DMC.

Let us also consider the DFT results reported here in the context of
previous predictions.  Dobson
\etal\cite{dobson2012calculation,white2008enhanced} and Misquitta
\etal\cite{misquitta2010dispersion} have previously identified key
assumptions in vdW approximations based on $1/r^6$ behavior. They
point out that such approximations (present in the dispersion
corrections and vdW-DFs) render these methods incapable of accounting
for non-additivity in polarizabilities, which are particularly
relevant for extended low-dimensional systems such as
CNTs. Nonetheless, DMC has been used to show in this study that
dispersion corrected methods perform relatively well for molecular
adsorption on a non-metallic CNT.

The neglect of non-additivity in polarizabilities is expected to be
important in metallic or small band gap
systems\cite{dobson2012calculation,white2008enhanced,misquitta2010dispersion}.
Therefore, it would be interesting to know if the performance of the
dispersion corrected methods holds for nanotubes with different
electronic properties. This would require more high-accuracy benchmark
calculations on a metallic CNT and an insulating nanotube such as a
BNNT. Although there are currently no DMC references for these
systems, we computed the adsorption energy of water inside and outside
a metallic CNT(6,6) and an insulating BNNT(10,0), with PBE+D3 and
vdW-DF. These nanotubes have similar diameters to the
CNT(10,0). Interestingly, we find that the same trends in energy are
exhibited in these nanotubes as for water on CNT(10,0), regardless of
the electronic properties of the nanotubes. The adsorption energy
outside the CNT(6,6) is $\sim-120$ meV with PBE+D3 and vdW-DF, whereas
inside the CNT(6,6), vdW-DF predicts a much stronger adsorption energy
($-420$ meV) compared to PBE+D3 ($-304$ meV).  The interaction
energies of water on BNNT(10,0) are 10-20 meV stronger than on the
CNTs, but vdW-DF still predicts a considerably larger interaction
energy for water inside the BNNT than PBE+D3. Thus, these calculations
indicate that the conclusions made here about the performance of
various \xc functionals likely apply to other systems.

\section{Conclusions}\label{conc}
Reference DMC interaction energies have been computed for water and
H$_2$ on the outside and inside of the zigzag CNT(10,0) and also for
H$_2$ on graphene. Adsorption of either water or H$_2$ inside this
nanotube is about three times larger than outside, suggesting that the
uptake of water and H$_2$ is possible in some sub-nanometre CNTs. With
regard to the wide-ranging experimental adsorption energies reported
for H$_2$ on carbon nanomaterials, the DMC reference energies for
H$_2$ corroborate that the adsorption energy is weak at around $-100$
meV or less. In addition, the adsorption energy of water on the CNT is
a factor of $\sim$2 larger than H$_2$ and thus, H$_2$ is less likely
to be adsorbed on a CNT in the presence of water.

Three water-carbon force fields were benchmarked against DMC,
including the widely used Werder \etal potential. Naturally, the
results are very sensitive to the parameters and underlying model, but
we find that for water on CNT(10,0) the force field model given by
P{\'e}rez-Hern{\'a}ndez and Schmidt predicts interaction energies in
good agreement with DMC. In contrast, a selection of widely used and
new \textit{xc} functionals considered here are unable to accurately
predict the interaction energies for these systems. Strikingly, there
is a clear distinction between dispersion corrected \textit{xc}
functionals - which only slightly overestimate the interaction
energies - and dispersion inclusive functionals. The latter strongly
over-bind molecules inside the CNT: up to twice as much. An analysis
of DFT energies indicates that the inaccuracy arises from medium-range
correlation, which seems to be poorly described by the dispersion
inclusive functionals.  These findings also hold for molecular
adsorption inside a metallic CNT and a BNNT, indicating that the error
from medium-range correlation is wide-spread and likely to manifest in
other systems. Indeed, benchmark studies of water on other
low-dimensional materials suggest they too lack consistent
accuracy\cite{Ma11a,al-hamdani2}.

Finally, we expect that the reference adsorption energies of water and
H$_2$ on CNTs established in this work, will help to understand and
interpret studies regarding bio-sensing, storage capacities, slip
lengths, and molecular transport in CNTs, among other applications.

\acknowledgments We are grateful for support from
University College London and Argonne National Laboratory (ANL)
through the Thomas Young Centre-ANL initiative. Some of the research
leading to these results has received funding from the European
Research Council under the European Union's Seventh Framework
Programme (FP/2007-2013) / ERC Grant Agreement number 616121
(HeteroIce project). A.M. is supported by the Royal Society through a
Wolfson Research Merit Award. This research also used resources of the
Argonne Leadership Computing Facility at Argonne National Laboratory,
which is supported by the Office of Science of the U.S. DOE under
contract DE-AC02-06CH11357. This research also used resources as part
of an INCITE project (awarded to D.A.)  at the Oak Ridge National
Laboratory (Rhea/Eos), which is supported by the Office of Science of
the U.S. Department of Energy (DOE) under Contract
No. DEAC05-00OR22725. In addition, we are grateful for computing
resources provided by the London Centre for Nanotechnology and
Research Computing at University College London.

\bibliographystyle{apsrev4-1}\bibliography{updated-lib}

\newpage

\setcounter{section}{0}

\renewcommand{\thesection}{S\arabic{section}}%

\setcounter{table}{0}

\renewcommand{\thetable}{S\arabic{table}}%

\setcounter{figure}{0}

\renewcommand{\thefigure}{S\arabic{figure}}%

\section*{Supporting Information}

\section{abstract}
Here we have included interaction energy curves for water outside
CNT(10,0) and a graphene substrate from density functional theory.  In
addition, we demonstrate the prominence of medium-range water/carbon
distances for water adsorbed inside CNTs with different
diameters. Finally, the structures used to compute benchmark
interaction energies are provided.

\section{Water on carbon nanomaterials}
\begin{figure}[ht]
\centering \includegraphics[width=0.65\textwidth]{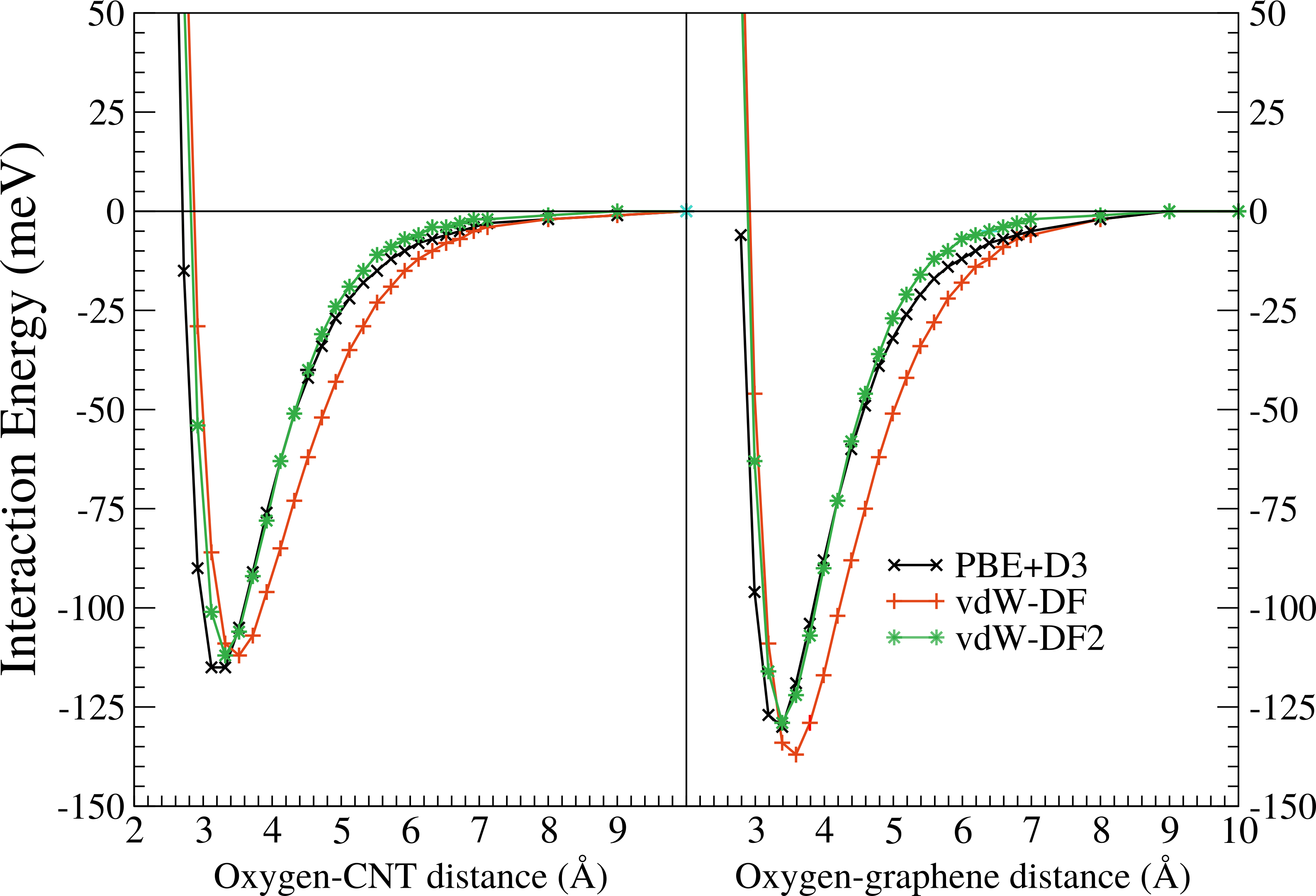}
\caption{Interaction energy curves for water outside the CNT (left)
  and graphene substrate (right) using PBE+D3, vdW-DF and
  vdW-DF2. Each point represents the interaction energy in meV from
  single point calculations of the water-substrate complex at a given
  height above the substrate.}\label{si-figure1}
\end{figure}

The interaction energy curve of water on graphene has previously been
computed with a few different \xc~ functionals, including vdW-DF2 and
vdW-DF\cite{Hamada2012,Ma11a}. We have also computed the interaction
energy curve for water on graphene in this study, as can be seen in
Fig.~\ref{si-figure1}.  The water/graphene interaction energy curves
computed here agree with previous results from Hamada\cite{Hamada2012}
with vdW-DF and vdW-DF2. In addition, the figure shows the interaction
energy curve for water outside the CNT(10,0) using PBE+D3, vdW-DF and
vdW-DF2 functionals. It can be seen that the water adsorption energy
from these functionals only vary by $\sim10$ meV outside the CNT.  The
water/graphene adsorption energy varies by the same amount. It can be
seen that the adsorption energy of water on graphene is higher than on
the CNT, by up to $\sim15$ meV according to these \xc~ functionals.
However, the interaction energy curves show that the vdW-DF
interaction energies are signficantly more than the vdW-DF2 or PBE+D3
energies at water-substrate distances between 3.5 to 6.0 \AA. The
vdW-DF2 and PBE+D3 interaction energy curves overlap for the most part
of this region. The results demonstrate that the medium-range vdW-DF
interactions are larger even for water outside the CNT and thus,
indicate a different balance of non-local correlation energy with
respect to other contributions, compared to the other functionals
considered.

\begin{figure}[ht]
\centering \includegraphics[width=0.65\textwidth]{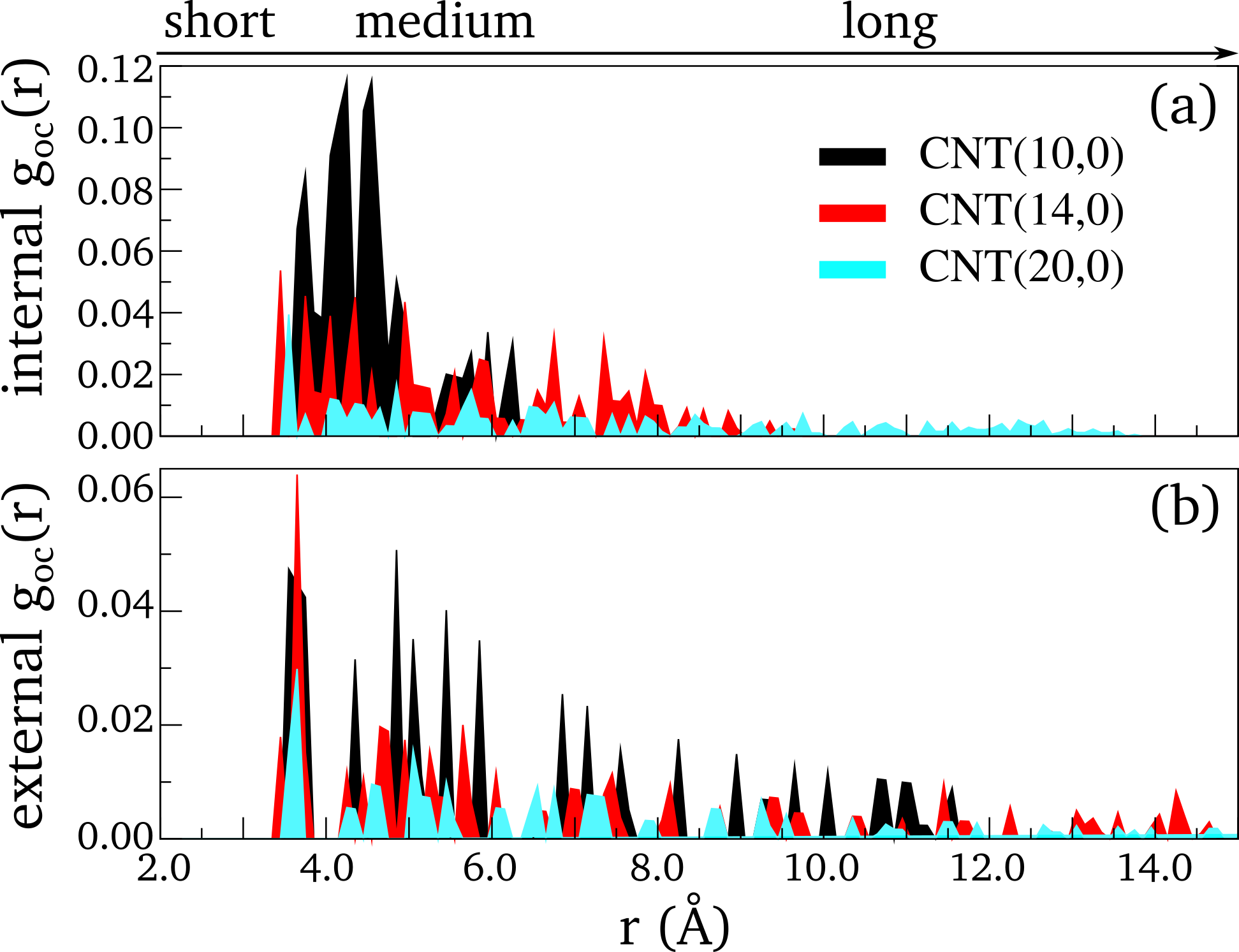}
\caption{Radial distribution functions of the oxygen-carbon bonds for
  water inside (top panel) and outside (bottom panel) CNTs with
  different diameters.}\label{si-figure2}
\end{figure}

In Fig.~\ref{si-figure2}, we show the radial distribution function for
oxygen-carbon inside and outside CNTs. The top panel of
Fig.~\ref{si-figure2} shows that the number of oxygen-carbon bonds for
water inside the CNTs increase in the medium-range of distances (3.5
to 6.0 \AA) as the CNT diameter decreases. On the other hand, the
number of oxygen-carbon bonds in the medium-range do not show a marked
variation with CNT diameter (see bottom panel of
Fig.~\ref{si-figure2}). Considering that dispersion inclusive
functionals appear to have larger errors in the medium-range
distances, we can see why they perform poorly for water inside
CNT(10,0), but come closer into agreement with dispersion corrected
functionals as the CNT diameter increases. As for larger CNTs, there
are less medium-range bonds, and hence a reduced error from
medium-range interactions.

The structures for H$_2$ and water inside and outside the CNT, and
H$_2$ on graphene, as computed with benchmark DMC, are provided in
separate xyz files.

\end{document}